# Implementing the draft Graph Query Language Standard

The Financial Benchmark

Malcolm Crowe
Emeritus Professor
University of the West of Scotland
United Kingdom
e-mail: malcolm.crowe@uws.ac.uk

Fritz Laux
Emeritus Professor
Reutlingen University
Germany
e-mail: fritz.laux@reutlingen-university.de

*Abstract*—The International Standards Organization (ISO) is developing a new standard for Graph Query Language, with a particular focus on graph patterns with repeating paths. The Linked Database Benchmark Council (LDBC) has developed benchmarks to test proposed implementations. Their Financial Benchmark includes a novel requirement for truncation of results. This paper presents an open-source implementation of the benchmark workloads and truncation.

## I. INTRODUCTION

The growth in the use of graph models has led to the development of standards including the publication of ISO 9075-16: Property Graph Queries (PGQ) [1], and the imminent emergence of a draft international standard for Graph Query Language (GQL) [2]. These developments draw on experience with commercial graph database products and envisage a clear convergence at the conceptual level between graph-based and relational database management, while GQL remains a separate standard. The principal novelty of GQL is its support for repeating graph patterns, which are useful in many applications including detection of fraud, analysis of supply chains, and cybersecurity [3].

Our previous work [4] has recommended the implementation of graph databases by extending the capabilities of a suitable Relational Database Management System (RDBMS) using metadata and additional built-in data types and syntax, particularly for the graph-oriented CREATE and MATCH statements, and has presented a working open-source solution that conforms to the usual requirements for RDBMS including transactions and security. In this paper we present an open-source RDBMS implementation, PyrrhoDB [5] that is able to perform graph creation and pattern matching including repeating patterns and also aligns well with the draft international GQL standard.

In particular, we will focus on the Financial Benchmark from the Linked Data Benchmark Council (LDBC), which explores the important use case of fraud detection and contains sample databases and illustrative workloads. The benchmark allows the performance of different implementations to be compared and introduces the new concept of truncation for managing the extent of searching, especially for historical data.

The benchmark envisages a database built to collect data on transfers between (possibly blocked) accounts, multiple ownership of accounts and relationships with and between companies, loan applications, guarantees, and remote operation of accounts (possibly using blocked or stolen devices), with a view to discovering and documenting criminal behavior including theft, fraud, and money laundering. The UML diagram is shown in Figure 1.

When graph databases contain event data accumulated over years, simply searching for a particular suspicious graph pattern can take an unreasonably long time. In extreme cases, where early detection is important (tight latency requirements), but nodes of interest have millions of edges to be investigated (power-law distribution of data) despite all available restrictions, it can become desirable to have a tunable mechanism to *truncate* the number of edges searched at each stage. The proposal in the benchmark is to maintain deterministic behavior by specifying a specific ordering to be used when the number of edges to be traversed exceeds a threshold. This threshold should be tunable on a per-query basis.

Naturally, the benchmark does not specify a mechanism for truncation. In this paper we offer an efficient implementation of this concept suitable for the direct, incremental, search algorithm in our open-source RDBMS.

The plan of this paper is to review the new implementation details in Section II. Section III presents an illustrative example, and Section IV provides some conclusions.

## II. IMPLEMENTATION DETAILS

We begin with a brief review of the graph pattern matching support in the standard, and the syntax definitions used in our relational database implementation. Further details are available in the references. Section B below discusses LDBC's truncation concept and the added syntax for this feature used in our implementation.

### A. Node and Edge Types

Our implementation of GQL using relational technology is fully described in [4] and [5]. Its database server accepts and directly implements both SQL and GQL source from the client, and its storage consists of the transaction log.

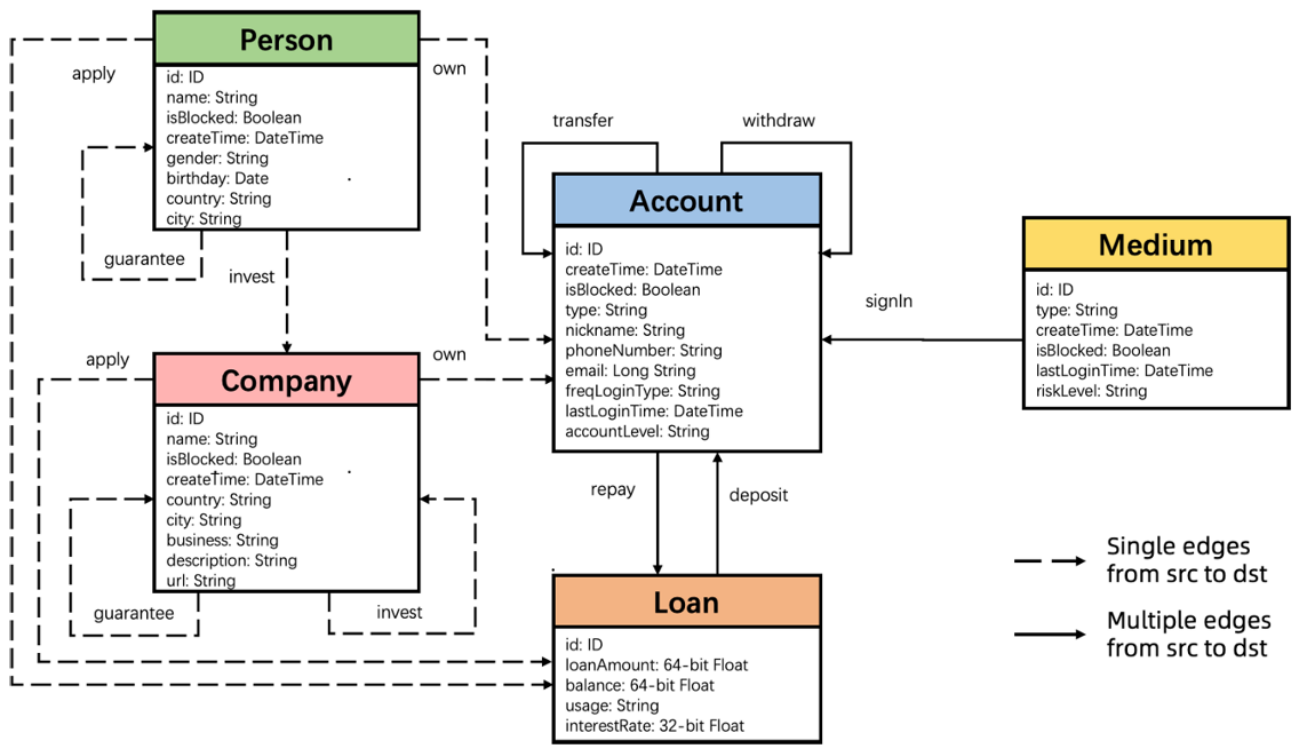

Figure 1. The LDBC FinBench data schema (from [6])

Specifically, GQL's CREATE statement is executed by a deep traversal of its graph patterns described while obeying the implied SQL DDL and DML actions, and the MATCH statement has its own query engine, which constructs a derived table resulting from a deep traversal of its graph patterns. Most MATCH statements have a dependent statement: for example, the GQL YIELD or RETURN statement, which provides expressions to construct results for the client. For example:

```
MATCH (P:Person{name:'Hatfield'})-[:own]
    ->(A:Account) return P.id,A.id
```

This example will give a table showing the id and account numbers of the persons called Hatfield. Unbound identifiers such as P and A above can be introduced at any point in the pattern, as shown in our next example, which also shows a MATCH statement without a dependent statement.

The MATCH statement allows the user to specify a graph fragment in queries instead of using joins. For example, with the scenario shown in Figure 1, the following query shows the details of all transfers in the database from any account owned by Hatfield:

```
MATCH (:Person{name:'Hatfield'})-[:own]->()
-[:transfer{amount:m,"timestamp":d}]->()
<-[:own]-(:person{name:r})
```

Figure 2 shows the result when this query is applied to the small LDBC financial benchmark database sf001.

The implementation begins by treating node and edge types are special kinds of SQL user defined types. Then each node or edge type corresponds to a base table in the relational database, whose rows are specific nodes and edges in the graph. Edges must identify two nodes: for directed

```
SQL> MATCH (:Person{name:'Hatfield'})-[:own]->()-[:transfer{amount:m,"timestamp":d}]->()
<-[:own]-(:person{name:r})
|----------------|----------------------|------------------|
|M               |D                     |R                 |
|----------------|----------------------|------------------|
|2977613.82      |07/10/2022 04:35:24   |Skundric          |
|6888877.75      |16/10/2022 03:43:21   |Hamahang          |
|989617.6        |26/10/2022 18:14:50   |Buchkin           |
|4024112.15      |27/10/2022 10:04:02   |Alfaro Siqueiros  |
|----------------|----------------------|------------------|
SQL>
```

Figure 2. A simple MATCH statement applied to the LDBC Financial Benchmark database sf001 (available from[6]).

edges these are called the source or leaving node and the destination or arriving node. The implementation constructs primary and foreign key indexes to support this structure.

The CREATE statement allows creation of nodes and associated edges in line using special tokens: nodes are enclosed in parentheses and edges join these using tokens -[ , ]-> , <-[ , ]-. One or more such patterns can be provided in a single CREATE statement. Within the parentheses or square brackets there is provision for an alias, a type label, and properties. The alias can be used to refer to the node or edge later in the same statement. Execution of the CREATE statement constructs new nodes and edges in the database with the given properties.

The MATCH statement allows retrieval of graph data by providing one or more patterns of nodes, edges and properties, similarly written to the CREATE statement, which is tested against the contents of matching database tables. A pattern will generally yield a table of bindings of new identifiers encountered in the pattern, which can be used in a dependent statement (e.g., a CREATE or RETURN statement). The RETURN statement can also contain aggregations whose scope is the entire binding table.

In addition to the simple graph patterns such as those in CREATE statements, MATCH statements can contain quantified path patterns (an example is given below), which match a sequence of nodes and edges in the database that traverses the path pattern a number of times that conforms to requirements in the quantifier such as ? (0 or 1), + (1 or more), * (0 or more) or {a,b} (at least a and not more than b). The rules provide for management of duplicate edges, nodes, or bindings.

In the resulting binding table, aliases that occur within such repeating patterns will have values that are arrays: one element for each traversal of the path pattern.

B.  *The LDBC Truncation concept*

In the financial benchmark specification [6], there is a concern that in selecting edges to follow from a given node (for example, traversing a set of transfers to or from an account) there may be hundreds or even millions of edges at each step, resulting in billions of cases to consider. It suggests a mechanism "to do truncation on the edges when traversing out from the current vertex", and to specify a sort order on such vertices to achieve consistency of results.

Since the traversal mechanism takes place inside the implementation of the MATCH statement, it makes sense to us to allow the truncation parameters to be specified as part of the creation of the MATCH statement, and we have constructed a syntax for this. The full syntax for Match in PyrrhoDB is shown in Figure 3. It includes:

```
MatchStatement = MATCH
    [Truncation] Match {',' Match}
    [WhereClause] [Statement] .

Truncation = TRUNCATING TruncationSpec
    {',' TruncationSpec} .

TruncationSpec = [EdgeType_id]
    ['(' OrderSpec {',' OrderSpec} ')'] '=' int .
```

The Truncation clause defines an upper bound for the number of edges to be traversed from a node in a step of the match process. The limit can be applied differently to specific edge types. Limits specified for supertypes of selected edges are also applied, as is the unnamed limit if present. It is explicit in the financial benchmark specification that the resulting truncation is performed within the execution of the database engine, and it is made deterministic by the specified ordering. There is an example in Figure 4 below.

The financial benchmark describes the truncation order as an enumeration and gives example values that are specific to the benchmark scenario, such as `TIMESTAMP_DESCENDING` and `AMOUNT_ASCENDING`. The syntax for OrderSpec is not shown here: in its simplest form it is a column name, but it can be a scalar expression optionally followed by ASC or DESC. Neither SQL nor GQL specifies a mechanism passing a parameter such as this to a stored procedure, but textual substitution is supported in prepared statements, which thus implement the notion of *general parameter* found in the GQL draft standard.

C.  *The Financial Benchmark Example*

Figure 4 shows the first complex read-only query in the Financial Benchmark. The node types involved are Medium

```
MatchStatement = MATCH [Truncation] Match {',' Match} [WhereClause] [Statement] .
Truncation = TRUNCATING TruncationSpec{',' TruncationSpec} .
TruncationSpec = [EdgeType_id] ['(' OrderSpec {',' OrderSpec} ')'] '=' int .
Match = (MatchMode [id '='] MatchNode) {'|' Match} .
MatchNode = '(' MatchItem ')' {(MatchEdge|MatchPath) MatchNode} .
MatchEdge = '-[' MatchItem '->' | '<-' MatchItem ']-' .
MatchItem = [id | Node_Value] [GraphLabel] [ Document | Where ] .
MatchPath = '[' Match ']' MatchQuantifier .
MatchQuantifier = '?' | '*' | '+' | '{' int , [int] '}' .
MatchMode = [TRAIL|ACYCLIC| SIMPLE] [SHORTEST |ALL|ANY] .
```

Figure 3.  PyrrhoDBMS's Match statement syntax [5].

and Account, and the only edge type is Transfer. The accompanying text in [6] reads: "Given an Account and a specified time window between startTime and endTime, find all the Account that is signed in by a blocked Medium and has fund transferred via edge1 by at most 3 steps. Note that all timestamps in the transfer trace must be in ascending order(only greater than). Return the id of the account, the distance from the account to given one, the id and type of the related medium."

To be relevant for this example, each link in a transfer chain must occur later than its predecessor, and this is why the timestamps are constrained to be in ascending order. To implement this, we define the following stored function that compares a given timestamp with the timestamp property of the last element of the given array:

```
create function later (a Transfer array, t timestamp)
returns boolean
    begin
        declare c int=cardinality(a);
        if (c=0) then
                return true
        else
                return a[c-1]."timestamp"<t
        end if
    end
```

The specification uses the SQL reserved word **timestamp** as a property name, so double quotes are needed on each occurrence of the name of this property (the occurrence of timestamp in the function heading declares the parameter **t** as having type **timestamp**).

Our implementation of the complex query described above reads as follows (parameters are in red, outputs in blue, internal identifiers in green):

```
MATCH
 truncating Transfer
   ("timestamp" truncationOrder)=truncationLimit
 trail p=(m:Medium{isBlocked:true})
   -[:signIn where "timestamp">startTime and
      "timestamp"<endTime]->
   (:Account{id:otherId})
      [()-[x:transfer
 where "timestamp" >startTime and "timestamp" <endTime
      and later(p.x,"timestamp")]->()]{1,3}
   (:Account{id:id1})
 return
   otherId,
   (cardinality(p)-3)/2 as accountDistance,
   m.id as mediumId,
   m.type as mediumType
 order by (accountDistance,otherId,mediumId)
```

Cardinality is an SQL function, and the cardinality of the path **p** is the total number of nodes and edges traversed: the formula here computes the account distance as the number of Transfer edges traversed.

The path identifier gives SQL code such as the above access to the binding table during and after construction, so that **p.x** above refers to the current value of the **x** column of the binding table, that is, before the new **x** edge is added to it. On the other hand, **p** also gives access to the path of nodes and edges, so that **p[i]** is the **i**th member of the path (a node or an edge), and the cardinality of **p** is the length of the path.

Despite the multiple joins implied and the repeated execution of the stored procedure, execution of this statement is commendably fast: on the sf0 sample database from LDBC, with the truncation defined as **transfer ("timestamp" desc)=10**, start time **timestamp'2022-01-01'**, end time **timestamp'2022-12-31'**, and id1 **4884435270860017215** it yields 3 rows in 7 seconds. Having identified the otherId accounts involved (here, 223491131508261941), an investigator can then investigate further.

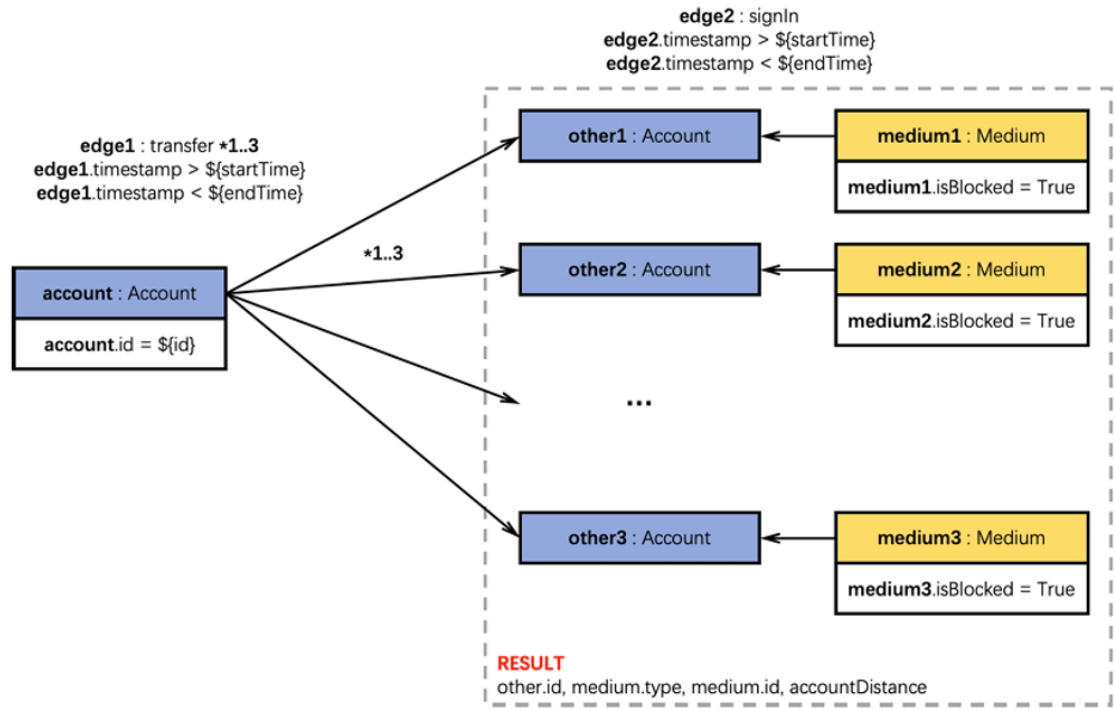

Figure 4.  A complex read-only query (from [6]).

### III. CONCLUSIONS AND FURTHER WORK

This step in our research into database technology was inspired by the LDBC Financial Benchmark [6], which suggested that truncation of graph pattern matching will often be a practical necessity for large graphs. We have proposed a general mechanism for search truncation, which in initial tests seems to be usable for searches in any property graph. With this in place, our prototype implementation is able to perform search efficiently even in large graphs.

As implementations of the draft international standard 39075 start to appear, there will be an opportunity to refine our proposal and compare it with other implementations of the benchmark.

### REFERENCES

[1] ISO 9075-16 Property Graph Queries (SQL/PGQ), International Standards Organisation, 2023.

[2] https://www.GQLStandards.org, October 4, 2023 – GQL status update [retrieved 18 October 2023].

[3] N. Francis et al.. A Researcher's Digest of GQL. 26th International Conference on Database Theory (ICDT 2023), Mar 2023, Ioannina, Greece. doi:10.4230/LIPIcs.ICDT.2023.1, pp. 1-22. https://hal.science/hal-04094449 [retrieved: 18 October 2023]

[4] M. Crowe and F. Laux, "Database Technology Evolution II: Graph Database Language", IARIA International Journal on Advances in Software, vol. 16 numbers 3 and 4, 2023, pp. 192-203, ISSN: 1942-2628.

[5] M. Crowe, PyrrhoV7alpha, https://github.com/MalcolmCrowe/ShareableDataStructures [retrieved: Dec 2023].

[6] Linked Data Benchmark Council: The LDBC Financial Benchmark (version 0.1.0), https://arxiv.org/pdf/2306.15975.pdf [retrieved Jan 2024].